\documentclass[a4paper,11pt]{article}
\usepackage{pos}
\renewcommand{\logo}{\relax}
\usepackage{braket, bbold}
\def \beq  {\begin{equation}}
\def \eeq  {\end{equation}}
\def \beqar {\begin{eqnarray}}
\def \eeqar {\end{eqnarray}}
\allowdisplaybreaks
\def\sqr#1#2{{\vcenter{\vbox{\hrule height.#2pt
\hbox{\vrule width.#2pt height#1pt \kern#1pt
\vrule width.#2pt}\hrule height.#2pt}}}}

\def\L {{\cal L}}

\def\Tr {{\rm Tr}}

\def\del {\partial}
\def\bdel{\bar{\partial}}
\def\a {\alpha}

\def\l {\lambda}

\def\A {{\cal A}}

\def\F {{\cal F}}

\def\H {{\cal H}}

\def \L {{\cal L}}
\def\M{{\cal M}}

\def\P {{\cal P}}

\def\half{\textstyle{1\over 2}}

\mathchardef\mhyphen="2D
\title{Entanglement Entropy \& Matter-Gravity Couplings for Fuzzy Geometry}
\author*[a]{V.P. Nair}
\affiliation[a]{City College of New York, CUNY\\
New York, NY 10031, USA}

\emailAdd{vpnair@ccny.cuny.edu}
\abstract{In this talk I discuss some features of the entanglement entropy for fuzzy geometry, focusing on its dependence on the background fields and the spin connection of the emergent continuous manifold in a large $N$ limit.
Using the Landau-Hall paradigm for fuzzy geometry, 
this is argued to be given by a generalized Chern-Simons 
form, making a point of connection with the thermodynamic view of gravity. Matter-gravity couplings are also considered in the same framework;
they naturally lead to certain specific nonminimal couplings involving powers of the curvature.
}

\FullConference{%
Corfu Summer Institute 2021 "School and Workshops on Elementary Particle Physics and Gravity"\\
  29 August - 9 October 2021\\
  Corfu, Greece
}

\begin{document}
\maketitle

\section{Introduction}

In this talk, I will present two results related to the use of fuzzy geometry
as the underlying structure for a theory of gravity.
This is based on the work published in \cite{nair1} and \cite{nair2}.
Let me begin by recalling that
 in fuzzy geometry we have an $N$-dimensional Hilbert space of states
$\H_N$, which maybe viewed as describing the (dynamical)
degrees of freedom pertaining to space itself.
Observables must be defined in terms of the algebra of matrices or linear transformations acting on $\H_N$; we will refer to this algebra as
${\cal M}_N$.
Space as a continuum is obtained as an approximation
in the limit $N \rightarrow \infty$.
This limit, however, can be ambiguous if we do not specify additional data beyond just the abstract structure of $\H_N$ itself. Such data usually take the form of a Dirac operator or Laplace operator, but the key point is that the choice of such an operator leads to a specification of the metric 
(and other geometric) data on the emergent space and hence the procedure for making this
choice (presumably implemented dynamically based on matter content) should be considered as what is meant by
a theory of gravity in the fuzzy context.
Rather than specifying a Laplace or Dirac operator at the level of
$\H_N$, one can think of the 
possible large $N$ limits as parametrized in terms of the gauge fields and spin connections in the final emergent continuous manifold.
In other words, we can take the starting data for fuzzy geometry
as $(\H_N, \M_N)$, along with a procedure for taking
large $N$ limits. This procedure can be stated in terms of a set of gauge fields and spin connections on the final emergent 
manifold.\footnote{It may be worth emphasizing
that the gauge fields we are talking about here are not the usual electroweak or chromodynamic gauge fields
of the standard model. We are talking about the gauge fields relevant to the definition of the fuzzy space itself. }
A theory of gravity is then a prescription for choosing one specific
large $N$ limit, which we may consider as ``optimal", the field equations for gravity are this optimization condition.
This is the setting for the results discussed below.

The first of the two results is about the entanglement entropy (EE) for a fuzzy space, i.e., entanglement pertaining to the degrees of freedom of 
space itself. In the usual way, here we are considering a division of space into two regions,
and looking for a (reduced) density matrix relevant for local observables in one
of these regions. The result for the entropy should thus follow from a suitable reduction of $\H_N$. 
The dominant term for the EE will be, as usual,
proportional to the area of the interface. One can even see that
there are some features reminiscent of the type ${\rm III}_1$
von Neumann algebra for the local observables \cite{nair1}.
But here my focus will be on the
dependence of this EE on the gauge fields and spin connection in the emergent continuum. We will argue that this dependence is given by a generalized Chern-Simons (CS) form
\cite{nair1}.
The interest in the question of how EE may be related to the
gauge fields and spin connection and how it can inform the issue of gravity for fuzzy spaces are due to the following observations. As has been known for many years, the
extremization of the Bekenstein-Hawking (BH) entropies for Rindler horizons of all accelerated observers in a spacetime leads to the Einstein equations of gravity in vacuum \cite{jacobson}.
Secondly, at least in some contexts, the BH entropy can also be viewed as an entanglement entropy \cite{RT}.
Finally for gravity in (2+1) dimensions (and for CS-type gravity in higher dimensions) the field equations correspond to the extremization of appropriate CS forms \cite{{grav3},{zan}}. We see that the connection between EE for fuzzy spaces and CS forms becomes relevant within this circle of known results on gravity.

Our second result is on the interaction of matter fields with the gauge fields and spin connection mentioned above. Again, to make our statement regarding this a little more precise, consider the situation where $\H_N$ is the space of sections of a power of the canonical line bundle on a compact K\"ahler space, i.e., arising from the geometric quantization of
the symplectic structure $\Omega = n \, \omega \equiv d a$, where $\omega $ is the K\"ahler two-form.
This shows clearly that $a$ (which is an Abelian gauge field defining
$\Omega$ and hence the structure of $\H_N$) should be one of the gauge fields of interest, relevant to defining the fuzzy space itself.
And more generally we can consider
$A = a + A'$ where we add some perturbation to the starting $a$,
without changing the topological class of $\Omega$.
We could also consider perturbations of the spin connection on this K\"ahler space. As we will see later, we could further include additional structures
corresponding to nonabelian gauge fields as well.
Let $S(A)$ denote the CS form of the gauge fields and the spin connection, we will make the specific nature of this term very explicit later. The second result I shall discuss is that the coupling of matter fields is then given by
$S(A+ {\mathbb A} )$ where the $U(1)$ field $A$ is shifted by
the Poincar\'e-Cartan one-form ${\mathbb A}$ for matter
 \cite{nair2}.\footnote{This is somewhat converse to the usual approach.
Instead of starting with a matter Lagrangian and ``gauging" it with, say, spin connection to obtain the coupling to gravity, here we start with
$S(A)$ given in terms of the spin connection and any other gauge fields
needed to define the fuzzy space and then shift $A$ to obtain the matter
coupling.}
This will give the action in the first order Hamiltonian formulation. In terms of the Lagrangian, the resulting action is
of the form
\beq
S = \int ({\rm polynomial~in~} R, F ) \times \L_{\rm matter}
\label{corfu1}
\eeq
This implies that in addition to the minimal coupling of mater fields to gauge fields and gravity, which is to be expected, there are specific nonminimal terms which is a polynomial in the curvatures ($R$) and gauge field strengths ($F$). The matter Lagrangian will have the usual covariant derivatives of fields and so on, but the density for integration over the volume has
a polynomial in terms of $F$, $R$, in addition to the usual
$\sqrt{\det g}$ factor.
It may be interesting to mention at this point that such couplings
have been used in some attempts to explain galactic rotation curves and similar phenomena usually attributed to dark matter \cite{dark}.

\section{Field dependence of entanglement entropy}

\subsection{How does field dependence arise for the EE?}

Turning to details, consider the complex projective space
$\mathbb{CP}^k$. We can carry out the geometric quantization of $\Omega = n \omega = d a$
in the holomorphic polarization.
The corresponding sections of the line bundle are holomorphic, they are the ``wave functions" which give a realization of the Hilbert space $\H_N$.
We can also view these wave functions as defining the lowest Landau level (LLL)
of a quantum Hall problem, with $\Omega$ as the background 
magnetic field. In other words, these wave functions are the lowest eigenstates of a Laplace operator, with covariant derivatives
defined with the connection $a$.
While the quantum Hall analogy is not necessary for our analysis, it is a useful picture
giving intuition about some of the arguments to follow \cite{{KN1},{KN2}}.
For example, in this picture, the state describing the fuzzy space 
is the fully occupied LLL with one fermion for each state.
If we consider the fermion field expanded as
\beq
\psi = \sum_s b_s \, u_s(x) + {\rm higher~ Landau~ levels}
\label{corfu2}
\eeq
where $b_s$ are fermion annihilation operators for the LLL and $u_s$ are the wave functions mentioned above,
fuzzy $\mathbb{CP}^k$ is the state $b_0^\dagger \, b_1^\dagger \ \cdots b_N^\dagger \ket{0}$ or, equivalently, it is described by the density matrix
\beq
\rho = b_0^\dagger \, b_1^\dagger \ \cdots b_N^\dagger \ket{0}
\bra{0} b_N  \cdots b_1 b_0
\label{corfu3}
\eeq
This is a pure state and if we reduce it to a subset of $b$'s (by tracing over the rest of them) we still obtain a pure state and hence no entropy. However, if we consider dividing $\mathbb{CP}^k$ into two regions, say, $D$ and its complement, then we do get a nontrivial entropy upon reducing
$\rho$ to $\rho_{\rm Red}$ relevant for local observables in $D$.
This entropy is given by
\beq
S_{\rm EE} = - \sum_s \left[ \lambda_s \log \lambda_s + (1-\lambda_s)
\log (1- \lambda_s) \right], \hskip .2in \lambda_s = \int_D u^*_s u_s
\label{corfu4}
\eeq
I will not go over the derivation of this formula since it has appeared in the literature before \cite{sierra} and has been used in related work \cite{{nair1},{kar1}} and
in the previous talk \cite{kar2}  to calculate the entanglement entropy for quantum Hall droplets.

A few comments might be appropriate at this point before we proceed to discuss the field dependence of the entropy.
First of all, we may note that the wave functions
$u_s$ have support everywhere on $\mathbb{CP}^k$, although they are exponentially small away from where $u^*_s u_s$ is a maximum.
This allows for a ``leakage" of $u^*_s u_s$ across any interface between
$D$ and its complement. This is the essence of how a nonzero $S_{\rm EE}$
can arise.
This is similar to what happens in relativistic quantum field theory where the
two-point function, for example, for a scalar field, is nonzero even for spacelike separations. This fact is, in turn, related to the nonfactorizability of the vacuum wave functional for the field, to the Reeh-Schlieder theorem and the result that
the reduced density matrices for local observables are KMS (Kubo-Martin-Schwinger) states with nonzero entropy.

Secondly, notice that the wave functions carry information about the
gauge field $a$ and the spin connection (and metric) on $\mathbb{CP}^k$.
In taking the large $N$ limit of an operator ${\hat F} $ on $\H_N$,
we construct the symbol which is a function (on $\mathbb{CP}^k$)
associated to it given by
\beq
F = \sum_{r, s} u_r\, F_{rs}\, u^*_s
\label{corfu4a}
\eeq
where $F_{r s} = \bra{r} {\hat F} \ket{s}$ are the matrix elements
of the operator ${\hat F}$.
Operator products  are then realized as star products of the symbols,
i.e.,
\beq
\sum_{r, s} u_r\, \bra{r} {\hat F} {\hat G} \ket{s} u^*_s
= \sum_{r, s, k} u_r\,  \bigl(F_{rk}  G_{ks} \bigr)\, u^*_s
= F * G
\label{corfu4b}
\eeq
In the large $N$ limit, the star product simplifies, becoming commutative,
so that the operator algebra $\M_N$ tends to the commutative algebra of pointwise multiplication of the symbols.
This is the procedure for
obtaining the continuum description. (See \cite{KN1} for explicit calculations for $\mathbb{CP}^k$.)
The expressions for the symbols and the star product will, of course,
depend on the
gauge field $a$ and the spin connection on the space
via the dependence of the wave functions $u_s$ on the same.
Each choice of the background field gives a particular large $N$ limit.
It is in this sense that I mentioned that the large $N$ limits are parametrized by the
background gauge fields.
More generally, one can consider the lowest eigenfunctions of the Laplacian
with more general background fields, including nonabelian gauge
fields
as well as general spin connections. So the data defining the wave functions and hence the large $N$ limits will then be these more general background fields.\footnote{Again, as mentioned in footnote 1, these are
not the gauge fields of
the standard model, these are part of the structure defining the
large $N$ limit of the fuzzy space itself.}
(As we consider different choices for
these background fields, it is important to keep $N$ the same, so that
the Hilbert space $\H_N$ has the same abstract structure.
This can be ensured by keeping all background fields in the same topological
class as defined by an index theorem; this will be explained below.)

\subsection{Moser's lemma and Chern-Simons term}

We can now turn to the crucial question we posed earlier.
How do $\lambda_s = \int_D u^*_s u_s$ depend
on the background gauge fields and the spin connection?
For analyzing this, we will again use $\mathbb{CP}^k$, starting with
$\Omega = n  d a$, and the standard spin connection for the
Fubini-Study metric, and making perturbations to both.
Towards the calculation of the change in $\lambda_s$, 
we define an ``occupancy matrix" $P$ for $\H_N$ as
\beq
(P)_{ab} =\begin{cases}
\delta_{ a b } \hskip .1in & a, b =0, 1, \cdots,  s-1\\
0& {\rm otherwise}\\
\end{cases}
\label{corfu5}
\eeq
(We are still considering the state (\ref{corfu3}) with all states occupied, $P$ is just an auxiliary quantity to help with the argument here.) 
As mentioned above, the
 large $N$ simplification 
of matrices on $\H_N$  is facilitated by using functions which are the symbols \cite{KN1}. From (\ref{corfu4a}), the symbol or the function corresponding to the matrix $P$ is
$(P)_{s-1} = \sum_{a=0}^{s-1} u^*_a u_a$.
Therefore we can write
\beq
u_s^* u_s = (P)_{s} - (P)_{s-1}
\label{corfu6}
\eeq
To obtain the background field dependence of this quantity,
we look at 
$W \equiv  \Tr (P A_0 )_s$, 
in terms of which
\beq
{\delta W \over \delta A_0} = (P)_{s} + {\rm *\mhyphen product~corrections}
\label{corfu7}
\eeq
Here $A_0$ is just a dummy variable used to define $W$; it
can be set to zero after the functional derivative is taken, although it may be viewed as the time-component of a $U(1)$ gauge field.
The point is that $W$ is easier to calculate. Once we have $W$, equations
(\ref{corfu7}), (\ref{corfu6}) will lead us to $u^*_s u_s$, including its dependence on the gauge fields and spin connections.

First let us focus on changing the background values only for the
$U(1)$ field which occurs in $\Omega$, i.e., just for the canonical one-form.
The symplectic form is thus given by
$\Omega =  (n \omega  + dA') = \Omega_0 + d A'$. We can now relate 
$\Tr (P A_0 )$ calculated with $\Omega_0 + d A'$ (i.e., calculated with
the wave functions corresponding to $\Omega_0 + d A'$) to
a trace calculated with $\Omega_0$ as
\beq
\Tr ( P A_0 )_{\Omega_0+ dA'} = \Tr (P {\A} )_{\Omega_0} = \int  (P)* {\A} = 
\int  (P)\, {\A} + \cdots
\label{corfu8}
\eeq
where ${\A}$ is to be determined in terms of $A'$ and $\Omega_0$.
The integral in (\ref{corfu8}) is over the whole manifold.
The point is that $\Omega_0$ and $\Omega_0 +dA'$ belong to the same topological class, so they can be related by a diffeomorphism, i.e., by a change of coordinates as $v \rightarrow v - w$. Here we are essentially using Moser's lemma familiar from classical mechanics. The condition we need is
\beq
\Omega_0 + d A' \Bigr]_{v-w} = \Omega_0\Bigr]_v, \hskip .3in
\A = A_0\Bigr]_{v-w}
\label{corfu8a}
\eeq
In terms of the one-form potentials, this becomes
\beq
a + A' \Bigr]_{v-w} - a\Bigr]_v = d f \approx 0
\label{corfu9}
\eeq
Here $f$ is any function on the manifold and $\approx$ indicates equality up to an exact form.
Taking $a$ (with $d a = \Omega_0$) as the term of the
zeroth order, and $A'$ as of the first order, we can solve for
$w$ as a series in the perturbation.
For the first  two orders, the condition (\ref{corfu9}) reduces to
\footnote{One can, of course, include higher orders, but these will suffice to illustrate the argument.}
\beqar
&&w_1^j \del_j a_i  + a_j \del_i w_1^j - A'_i \approx 0\nonumber\\
&&w_2^j \del_j a_i + a_j \del_i w_2^j + w_1^j \del_j A'_i + A'_j \del_i w_1^j
- {1\over 2} w_1^k w_1^l \del_k \del_l a_i -
w_1^k \del_k a_j \del_i w_1^j \approx 0
\label{corfu10}
\eeqar
By adding certain exact one-forms, (since we only have
the weak equality $\approx$),
these equations can be solved as
\beqar
&&w_1^j = - (\Omega_0^{-1})^{jk} A'_k\nonumber\\
&&w_2^j = - (\Omega_0^{-1})^{jk} \left[
(\del_k A'_l - \del_l A'_k) w_1^l + {\textstyle{1\over 2}} w_1^m w_1^n \del_m (\Omega_0)_{nk}
+ {1\over 2} (w_1^m \del_k w_1^n) (\Omega_0)_{mn}\right]
\label{corfu11}
\eeqar
We can now calculate $\A = A_0\Bigr]_{v-w}$ as
\beqar
\A &=& A_0 - (\Omega_0)^{-1 ij} A'_i \del_j A_0 
+ (\Omega_0)^{-1 i j} \Bigl[\left(
{\textstyle{1\over 2}} A'_j \del_k A'_i 
- A'_j \del_i A'_k \right) (\Omega_0)^{-1 k l} \del_l A_0\nonumber\\
&&\hskip .2in
- {\textstyle{1\over 2}} A'_j A'_k \del_i \left( (\Omega_0)^{-1 kl}\del_l A_0
\right)\Bigr] + \cdots
\label{corfu11a}
\eeqar
We can now multiply this expression by $P$ and integrate over the whole manifold to get
\beqar
W &=&
\int A_0 \left[ P + d \left( P k \Omega_0^{k-1} A' ) +
{P\over 2} k (k-1) \Omega_0^{k-2} d A' \, A' + \cdots\right)\right]
\nonumber\\
&&\hskip .2in + {k\over 2}\int P\, \Omega_0^{k-1}  d\bigl[ A' ( \Omega_0)^{-1ij} A'_i \del_j A_0\bigr]
+ \cdots
\label{corfu11b}
\eeqar  
The factors of $k$, $(k-1)$, etc. arise from writing the expression in terms of
differential forms. Again using integration by parts and rearranging
factors, we can write the last term in terms of $A_0$, rather than its derivative, to obtain
\beqar
W &=&
\int A_0 \left[ P + d \left( P k \Omega_0^{k-1} A' ) +
{P\over 2} k (k-1) \Omega_0^{k-2} d A' \, A' + \cdots\right)\right]
\nonumber\\
&&\hskip .2in - {1\over 2}\int A_0 d \left( 
 \Omega_0^{k-1}  A' ( \Omega_0)^{-1ij} \del_i P\,A'_j\right)
+ \cdots
 \label{corfu12}
 \eeqar
Taking the functional derivative with respect to $A_0$ we get
 \beqar
 \Omega_0^k P\bigr]_{a+ A'} &=& \Omega_0^k P + d \left( P k \Omega_0^{k-1} A' ) +
{P\over 2} k (k-1) \Omega_0^{k-2} d A' \, A' + \cdots\right)\nonumber\\
&&\hskip .2in - {1\over 2} d \left( 
 \Omega_0^{k-1}  A' ( \Omega_0)^{-1ij} \del_i P\,A'_j\right)
 \nonumber\\
 &=& \Omega_0^k P + d \bigl(P Q(a+A', a)\bigr) - {1\over 2} d \left( Q(a+A', a) 
 ( \Omega_0)^{-1ij} \del_i P\,A'_j\right)
 \label{corfu12a}
 \eeqar
 where
 \beq
 Q (a+ A', a) = k \int_0^1 dt ( d a + t\, dA' )^{k-1} \, A'
 \label{corfu12b}
 \eeq
 The left hand side of (\ref{corfu12a}) corresponds to calculations with
 $a+A'$, while the $P$'s on the right hand side are in terms of
 $u^*u$ with just $a$, i.e., with $A' = 0$.
 The whole expression is linear in $P$, so we can easily take the difference
 of such terms with $P_s$ and $P_{s-1}$, to obtain the field-dependent corrections
 to $u^*_s u_s$.
This result is
\beqar
\Omega_0^k \left[ (u_s^* u_s)_{A'_i \neq 0 } - (u_s^* u_s)_{A'_i = 0 }\right] 
&=& d \left[ (u_s^* u_s)_{A'_i = 0} \, Q (a+A', a)\right]\nonumber\\
&&- {k \over 2} d \left[ \Omega_0^{k-1} A' (\Omega_0^{-1})^{ij} \del_i (u_s^* u_s)_{A'=0} \, A'_j
\right] +\cdots
\label{corfu13}
\eeqar
 (For more details on these calculations using the idea of Moser's lemma, see \cite{{nair1},{nair3}}.
These calculations can also be done using the symbols and star products, see \cite{{KN1},{KN2}}.)
We can now integrate this result over $D$ to obtain the change in $\lambda_s$ 
due to $a \rightarrow a + A'$ as
\beq
\lambda_s\Bigr]_{A'_i \neq 0} - \lambda_s\Bigr]_{A'_i = 0} =
\oint_{\del D} Q (a+ A', a) \, (u_s^* u_s) - {1\over 2}\oint_{\del D}  \left[ Q (a+ A', a) \, (\Omega_0^{-1})^{ij} \del_i (u_s^* u_s)\, A'_j \right] + \cdots
\label{corfu14}
\eeq
In the integrands on the right hand side, $u^*_s u_s$ and other quantities are evaluated on $\del D$.

The quantity $Q (a+ A', a)$ which is given in
(\ref{corfu12b}) and appears in this formula is a generalized
Chern-Simons form connecting two nonzero 
gauge potentials {$A_1 = a$}, {$A_2= a+ A'$}.
If {$\P(F)$} is an invariant polynomial (such as an index density) expressed
as the symmetrized trace of a product of {$k$}
{$F$}'s, we define $Q(A_2, A_1)$ by
\beq
Q(A_2, A_1) =  k \int_0^1 dt~ \P (A_2 - A_1, F_t, F_t, \cdots, F_t),
\hskip .2in A_t = A_1 + t (A_2 -A_1)
\label{corfu15}
\eeq
This leads to
$\P (F_2) - \P(F_1) = d \, Q(A_2, A_1)$. Thus up to an exact form,
$Q(A_2, A_1)$ is the difference of two CS terms, but equation
(\ref{corfu15}) gives a specific formula for the extra exact form.
Such generalized CS forms are important in defining the anomaly with nontrivial background fields, see \cite{MSZ}.
For our simple Abelian case, choosing $\P (F) = F^k$, $A_2 = a +A'$,
 $A_1 = a$,
\beq
Q (a+ A', a) =   k \int_0^1 dt ( d a + t\, dA' )^{k-1} \, A' 
\label{corfu16}
\eeq
which agrees with (\ref{corfu12b}).
Upon comparison with (\ref{corfu14}), we see that the first correction to $\lambda_s$ is indeed proportional to this
generalized CS form. The second term in (\ref{corfu14}), which has additional
$\Omega_0^{-1}$, is subdominant at large $N$.
 Since $\delta S_{\rm EE}  = -\sum_s \delta \lambda_s \log ( \lambda_s/(1-\lambda_s))$, we can write
 \beq
 \delta S_{\rm EE} = \oint Q(a+A', a) \sum_s C_s \, (u^*_s u_s)_{\del D} 
 + \cdots, \hskip .2in C_s = \log (\lambda_s/(1-\lambda_s))
 \label{corfu16a}
 \eeq
Restating, our conclusion from this analysis is:
\begin{center}
\parbox{.8\textwidth}{
The leading field-dependent correction to the entanglement entropy is
proportional to the generalized Chern-Simons term $Q(a+ A', a)$.}
\end{center}

\subsection{Higher dimensions, the Dolbeault index}

This result can be generalized to include nonabelian gauge fields
and arbitrary gravitational backgrounds.
The wave functions for the LLL are holomorphic, they belong to the kernel of 
$\bdel$-operator and so the number of states 
is given by the Dolbeault index theorem \cite{KNindex}.
Since the bulk part of {$\Tr\, (P A_0)$} is related to the number of states,
it can also be calculated directly using the Dolbeault index
density.
The theorem states that the index is given as
\beq
{\rm Index} (\bdel_V) = \int {\rm td}(T_cK) \wedge {\rm ch}(V)
\label{corfu17}
\eeq
where ${\rm td}(T_cK)$ is the Todd class on the complex tangent space of $K$
and ${\rm ch}(V)$ is the Chern character of the vector bundle $V$.
The vector bundle refers to the fact that we can have 
gauge fields in addition to the gravitational fields.
The Todd class is given by
\beqar
{\rm td} &=& \prod_i {x_i \over 1- e^{-x_i}}\nonumber\\
&=& 1 + {1\over 2} \, c_1 +{1\over 12} ( c_1^2 + c_2) + {1\over 24} c_1\, c_2
+ {1\over 720} ( - c_4 + c_1\,c_3 + 3 \, c_2^2 + 4\, c_1^2 \,c_2 - c_1^4) + \cdots
\label{corfu18}
\eeqar
where the first line gives the formula in terms of the splitting principle and the second gives the expansion for low dimensions in terms of the Chern classes
$c_i$, which, for 
any vector bundle with curvature ${\cal F}$, 
 are defined by
\beq
\det \left( 1 + {i \, {\cal F} \over 2 \pi} \,t\right) = \sum_i c_i \, t^i
\label{corfu19}
\eeq
The Chern character is defined by
\beq
{\rm ch}(V) = \Tr \left( e^{i {\cal F} /2 \pi} \right) = {\rm dim}\,V + \Tr ~{{i {\cal F}} \over {2\pi}} + { 1 \over 2!} \Tr~ {{i{\cal F} \wedge i {\cal F}} \over {(2\pi)^2}} + \cdots
\label{corfu20}
\eeq 
With these formulae, one can see that the index, for low dimensions, is
contained in the expansion
\beqar
{\rm Index} (\bdel_V)
\!\! &=&\!\! \int {\rm dim} V\, \Tr \left({i R \over 4\pi}\right)  + \Tr \left( {i F \over 2 \pi}\right)
+ {{\rm dim}V \over 12} ( c_1^2 + c_2) + {1\over 2} \Tr \left( {i F \over 2 \pi} \right)^2
+ \cdots\nonumber\\
c_1 \!\!&=&\!\!\Tr {i R \over 2 \pi} , \hskip .2in
c_2 = {1\over 2} \left[ \left( \Tr {i R \over 2 \pi}\right)^2 - \Tr \left({i R \over 2 \pi}\right)^2
\right]
\label{corfu21}
\eeqar
Taking this index density as the invariant polynomial, we can define
a corresponding generalized Chern-Simons term by
the same formula as before, i.e.,
\beq
Q(A_2, A_1)= k \int_0^1 dt~ \P (A_2 -A_1, \F_t, \F_t, \cdots, \F_t),
\hskip .2in A_t = A_1 + t (A_2 -A_1)
\label{corfu22}
\eeq
where $\F$ can be $F$ or $R$, with $A$ referring to a gauge potential or the spin connection, respectively. Again, this definition (\ref{corfu22}) is 
consistent with
$\P (\F_2) - \P(\F_1) = d \, Q(A_2, A_1)$.  The fields $A_2$ and $A_1$ belong to the same topological class in the sense that the integrated index is the same for both, and one can continuously connect $A_2$ to $A_1$
as in $A_t = A_1 + t (A_2 -A_1)$.
This is the meaning of the phrase ``without changing the topological
class of $\Omega$" in the introduction.
The rest of the argument for the field-dependence of $\lambda_s$ is similar to the Abelian case. (For more details, see \cite{nair1}.)
Equation (\ref{corfu14}) still holds with $Q(a+A, a)$ replaced by the
generalized CS form for the Dolbeault index given in (\ref{corfu22}).
The formula for the leading correction to the entropy is the same as
(\ref{corfu16a}), with $Q$ corresponding to the Dolbeault index density.
So we can now restate our result more generally as:
\vskip .1in
\begin{center}
\parbox{.8\textwidth}{
The leading field-dependent correction to the entanglement entropy is
proportional to the generalized Chern-Simons term $Q(a+ A', a)$ associated to the Dolbeault index density.}
\end{center}
This completes my statement of the first result mentioned in the introduction.

\section{The condensed matter perspective}

Although this meeting is on particle physics and gravity, it may be interesting to change hats for a quick aside and view our result from a condensed matter perspective. We are considering the Hall state where all one-particle 
lowest Landau level (LLL) states are filled, so the density matrix (\ref{corfu3}) corresponds to
the $\nu =1 $ integer quantum Hall state.
The fermions in the LLL may be viewed as a droplet of a fluid, which is effectively incompressible because of the Pauli exclusion principle.
For the same reason, the many-particle state is highly correlated.
For simplicity, let me consider the two-dimensional case, i.e.,
$\mathbb{CP}^1 \sim S^2$ or its flat limit as we take the radius to be large. 
(The generalization to higher dimensional quantum Hall systems
is straightforward.)
The symplectic form $\Omega_0$ is the magnetic field, $a$ being the
electromagnetic vector potential.
The responses of the state to variations of this field (i.e., under
$a \rightarrow a+ A'$) and to variations of the spin connection are related to the electrical conductivity and the Hall
viscosity, respectively. These are clearly quantities of physical interest.
In the same spirit, one can ask about the entanglement entropy which can arise when we restrict attention to observables defined locally in some region of the droplet.
As is to be expected, there will be a formally divergent constant term proportional to the phase volume \cite{kar2}, but the dependence on the background fields, from our result, is of the form
\beq
\delta S_{\rm EE} \sim {1\over 2 \pi} 
\oint_{\del D} \left[ A -a + {1\over 2} (\alpha - \alpha_0 )\right] \sum_s C_s  (u_s^* u_s)_{\del D}
\label{corfu22a}
\eeq
where $a$ corresponds to the starting constant magnetic field, $A- a = A'$ gives the perturbation to it, and likewise, $\alpha_0$ is the spin connection for
$\mathbb{CP}^1$ with the Fubini-Study metric and $\alpha - \alpha_0$ is the
perturbation to it. The EE has been argued to be useful in characterizing topological phases of matter, and so, in this context, I expect
the result (\ref{corfu22a}) should be of interest.

\section{Coupling matter fields}

We now return to the main topic and the second result mentioned in the beginning.
Again, we will consider the manifold $\mathbb{CP}^k$,
with possible perturbations to the gauge fields and spin connection 
within the same topological class as before.
Consider now a matter system with the degrees of freedom
described by a set of variables
{$\{ q_\lambda \}$} and conjugate variables 
{$\{ p_\lambda \}$}. The
relevant physical quantity for the quantum dynamics of these degrees of freedom is the transformation kernel, which, for an infinitesimal
change $\epsilon$ of time, is
\beq
\bra{q'} e^{- i H \epsilon} \ket{q} = \int [dp] \, \exp\left[ i p_\l (q'_\l - q_\l ) - H(p,q) \, \epsilon \right]
\label{corfu23}
\eeq
The variables $q_\lambda$ may be viewed as the coefficients in the mode expansion of an $N \times N$ matrix ${\hat q}$ with matrix
elements
\beq
{\hat q}_{ij} = \sum_\l q_\l \, (T_\l )_{ij}, \hskip .3in
i, j = 1, 2, \cdots, N,
\label{corfu24}
\eeq
where $\{ T_\lambda \}$ form an orthonormal basis for matrices acting on
$\H_N$. (This is how the matter system is placed in the fuzzy space.)
 As in the case of $P$, we can rewrite the $\{ T_\lambda \}$ in terms of 
functions which are symbols associated to the matrices.
In the present situation it is slightly easier to use the so-called contravariant
symbols which form the basis for the Berezin-Toeplitz quantization
procedure \cite{{BTquant},{nairBT}}. The contravariant symbol  $\phi$ associated to
${\hat q}_{ij}$ is defined by \footnote{The contravariant symbol is a classical function which, upon ``quantization" according to the formula (\ref{corfu25}), leads to the quantum version, i.e., to the matrix
${\hat q}$. The symbol we used in (\ref{corfu6}) started with the matrix and
obtained a classical function from it. That function is referred to as the covariant symbol. In either case, the definitions are made using the
wave functions $u_i$.}
\beq
{\hat q}_{ij} = \int d\mu \, u_i^* \, \phi \, u_j 
\label{corfu25}
\eeq
We can now convert the set of variables {$\{ q_\lambda , p_\lambda \}$}
to symbols and
products of them to star products; for example, we can write
\beqar
\sum_\lambda p_\l p_\l &=& \sum_{\l, \l'} \Tr ( p_\l T_\l ) (p_{\l'} T_{\l'} ) 
= \sum_{i,j} \int d\mu d\mu'\,\bigl[ u_i^*(z)  \Pi (z) u_j (z)\bigr]
 \, \bigl[u^*_j(z') \Pi (z') 
u_i(z')\bigr] \nonumber\\
&\equiv&
\int d \mu~\sum_i  u_i^* (\Pi* \Pi ) u_i
= \int d\mu\, \rho~ \Pi* \Pi
\label{corfu26}
\eeqar
The transition from the first line to the second line of this equation uses the definition of the star product for the contravariant symbols.
Using this equation, and similar formulae for other terms in (\ref{corfu23}),
the transformation kernel can be converted to 
the path integral
\beqar
Z &=& {\cal N} \int [ D \Pi\, D\phi ]\, \exp\left( i \int dt\,d\mu\, {\rho}~ \left[\Pi* {\dot \phi}  - H (\Pi, \phi )\right]\right)\nonumber\\
&=& {\cal N} \int [D \Pi\, D\phi]\, \exp\left( - \int d\mu\, {\rho}~{\mathbb{A}}
\right)
\label{corfu27}
\eeqar
where $\mathbb{A}$ is given by
\beq
\mathbb{A} = -i \left( \Pi* {\dot \phi}  - H (\Pi, \phi ) \right)\, dt
\label{corfu28}
\eeq
$\mathbb{A}$ is the Poincar\'e-Cartan one-form for the dynamics of
the variables $\{ q_\lambda, p_\lambda\}$, i.e., for matter fields.
(We have also generalized (\ref{corfu23}) to a finite change of time.)

In the path integral (\ref{corfu27}) and in the definition
of $\mathbb{A}$ in (\ref{corfu28}), all the products with
$\phi$ and the conjugate variable $\Pi$ involve star-products.
The Hamiltonian will contain such products as well,
among the fields and their conjugates. As an example, let
$T_\alpha$ be the basis matrices obeying the
$SU(k+1)$ algebra in the $N\times N$ representation; they form a subset of the full basis $\{ T_\lambda \}$ and commutators with $T_\alpha$ serve as
derivatives and local rotations.
Then we can consider
a Hamiltonian of the form
\beq
H = {1\over 2} \Tr \left[ {\hat \Pi} \,{\hat \Pi}  + \beta_1\, [T_\a, {\hat q}]\, [ T_\a, {\hat q}] + m_0^2 {\hat q} {\hat q}
\right] + g_0\, \Tr ( {\hat q}^4)
\label{corfu29}
\eeq
where
$\beta_1, m_0, g_0$ are constants. This Hamiltonian, if we evaluate the
trace using the expansion (\ref{corfu24}) (and a similar one for
${\hat \Pi}$) is expressed entirely in terms of $\{q_\lambda, p_\lambda\}$.
Using symbols and star products according to (\ref{corfu25}), (\ref{corfu26}),
$H$ in (\ref{corfu29}) becomes
\beq
H (\Pi, \phi)  = \int d\mu\, \rho~ \left[ {1\over 2} \left( \Pi * \Pi + \alpha_1 \,(\nabla_\a\phi)* (\nabla_\a\phi) 
+ m_0^2 \phi *\phi \right) + g_0\, \phi* \phi* \phi* \phi \right]
\label{corfu30}
\eeq
Going back to (\ref{corfu27}), the key point to note is the presence of
the factor
${\rho}$ which gives the density of states.
As I mentioned earlier, the number of states is given by the Dolbeault index theorem, so we can identify the density $\rho$ with the Dolbeault index density (up to terms which are total derivatives and hence vanish upon integration).
In the case of the entanglement entropy, we considered the difference of this index density for two different connections, which led to the $(2k-1)$-form $Q(a+A', a)$, appropriate for integration over
the $(2k-1)$-dimensional interface between $D$ and its complement.
It is also possible to ``integrate up" from the index density
to define a $(2k +1)$-form on
$\mathbb{CP}^k \times \mathbb{R}$, $\mathbb{R}$ being the time direction.
This CS $(2k+1)$-form is designed to have the property that upon variation with respect to $A_0$ (the time-component of
$U(1)$ field $A$) it will give the Dolbeault index density. Explicitly this CS form is given by \cite{KNindex}
\beq
S_{\rm eff} =
 \int \Bigl[ {\rm td}(T_c K) \wedge \sum_p  (CS)_{2 p+1} ( A)\Bigr]_{2 k+1}
+ 2 \pi\int \Omega^{\rm grav}_{2k+1}
\label{corfu31}
\eeq
where $(CS)_{2 p+1}$ is the usual Chern-Simons $(2p +1)$-form
for gauge fields and $\Omega_{2k+1}^{\rm grav}$ is defined by
\beq
{\rm dim}\,V [{\rm td} (T_c K)]_{2k+2} = d\, \Omega_{2 k+1}^{\rm grav}\label{corfu32}
\eeq
$S_{\rm eff}$ in (\ref{corfu31}) depends on the gauge fields $A$,
spin connection $\alpha$.
Because it is the integral of a differential form,
$S_{\rm eff}$ in (\ref{corfu31}) can have, at most, one power of
the time-component of the Abelian part of
$A$, i.e. with $A_0 dt$.
We can write (\ref{corfu31}) as
\beq
S_{\rm eff} = \int \rho \, A_0\, dt + S^{(0)}_{\rm eff}
\label{corfu32a}
\eeq
where $S^{(0)}_{\rm eff}$ is independent of $A_0$.
The coefficient of $A_0$, by construction, is the index density $\rho$,
which is also the field-dependent generalization of
$\sum_i u^*_i u_i$.
Since Poincar\'e-Cartan one-form $\mathbb{A}$ has only the time-component, we see that
we can also write
\beq
S_{\rm eff} (A+ \mathbb{A}, \alpha) = 
\int \rho\, \mathbb{A}  + S_{\rm eff} (A, \alpha )
\label{corfu32b}
\eeq
Going back to (\ref{corfu27}), we see that we can write the exponent in
as $S_{\rm eff} (A+ \mathbb{A}, \alpha)  - S_{\rm eff} (A, \alpha) $.
Rather than subtracting out $S_{\rm eff} (A, \alpha)$, we 
will keep this in the path integral, as it does correspond to the
dynamics of the background fields themselves
\cite{nair4}.
This finally brings us to our second result:
\vskip .2in
\begin{center}
\parbox{.8\textwidth}{
The path integral describing the dynamics of matter fields coupled to 
gauge and gravitational fields defining the large $N$ limit for fuzzy spaces
is given by
\beq
Z = {\cal N} \int [D \Pi \, D \phi] \, \exp \Bigl( i S_{\rm eff} (A+ \mathbb{A}) 
\Bigr)
\label{corfu33}
\eeq
where $\mathbb{A}$ is the Poincar\'e-Cartan form for the matter dynamics
and $S_{\rm eff}$ is as given in (\ref{corfu31}).}
\end{center}

As an example, in four dimensions, the matter coupling takes the form
\beq
S_{\rm matter} = {1\over 32\pi^2}\int {(i \mathbb{A} )}\,
\Bigl[ {\rm dim}V \Bigl(F_{\mu\nu} F_{\alpha \beta} 
+ {1\over 24} R^{ab}_{\mu\nu} R^{ab}_{\alpha\beta}  \Bigr)
+ \Tr (t_a t_b) {\bar F}^a_{\mu\nu} {\bar F}^b_{\alpha\beta})\Bigr] dx^\mu\cdots dx^\beta
\label{corfu34}
\eeq
where 
$F = (-i )\, {\half} F_{\mu\nu} dx^\mu \,dx^\nu $ is the Abelian gauge field,
and ${\bar F} =  (-i t_a)\,{\half} F^a_{\mu\nu} dx^\mu \,dx^\nu$
is the nonabelian part of the background gauge fields, see footnote 1 for a
clarifying remark on this.
$R^{ab}_{\mu\nu}$ is the Riemann curvature tensor.
Thus $\mathbb{A}$ is multiplied by a polynomial of fields and curvatures 
whose form is determined
by the Dolbeault index.
If the integration  over the ``momenta" $\Pi$ is done in (\ref{corfu33}), one can write it in terms of a spacetime action. In this case the matter part takes the form mentioned in equation (\ref{corfu1}).

\bigskip
I thank the organizers of CORFU21 for the invitation and opportunity to 
present these results and for the wonderful
hospitality.
This work was supported in part by the U.S. National Science Foundation Grants No. PHY-2112729 and No. PHY-1820271.

%%%%%%%%%%%%%%%%%%%%%%%%%%%%%%%%%%%%%%%%%%%%%%%%

%%%%%%%%%%%%%%%%%%%%%%%%%%%%%%%%%%%%%%%%%%%%%%%%%%%%%%%%%%%%%%%%%%%%%%%%%%%%%%%%%%%%%%%%%%%%%%%%

\begin{thebibliography}{99}
%%%%%%%%%%%%%%%%%%%%%%%%%%%%%%%%%%%%%%%%%%%%%%%%
\bibitem{nair1} V.P. Nair, Phys. Rev. {\bf D 101}, 125021 (2020).

\bibitem{nair2} V.P. Nair, Phys. Rev. {\bf D 102}, 105008 (2020).

\bibitem{jacobson} See for example,
T. Jacobson, Phys. Rev. Lett. {\bf 75}, 1260 (1995).

\bibitem{RT} See for example,
S. Ryu and T. Takayanagi, Phys. Rev. Lett.~{\bf 96}, 181602 (2006);
JHEP~0608:045 (2006). (This is within the holographic framework.)

\bibitem{grav3} A. Ach\'ucarro and P. Townsend, 
Phys. Lett. {\bf B 180}, 89 (1986);
E. Witten, Nucl. Phys. {\bf B 311}, 46 (1988).

\bibitem{zan} For a recent general review, see
J. Zanelli, arXiv:0502193[hep-th].

\bibitem{dark} This is reviewed in T. Harko and F.S.N. Lobo,
Galaxies {\bf 2}, 410 (2014).

\bibitem{KN1} For the use of the symbol and the star product in this context, see D. Karabali and V.P. Nair, Nucl. Phys. { \bf B 641}, 533 (2002);
Nucl. Phys. { \bf B 679}, 427 (2004);
Nucl. Phys. { \bf B 697}, 513 (2004);
D. Karabali,  Nucl. Phys.~{ \bf B 726}, 407 (2005); Nucl. Phys.~{ \bf B 750}, 265 (2006).

\bibitem{KN2} D. Karabali and V.P. Nair, J. Phys. A Math. Gen. {\bf 39}, 12735 (2006);
D. Karabali, V.P. Nair and R. Randjbar-Daemi,
in {\it From  Fields to Strings: Circumnavigating Theoretical Physics},
Ian Kogan Memorial Collection, M. Shifman, A. Vainshtein and J. Wheater (eds.),
World Scientific, 2004; p. 831-876 and references therein.

\bibitem{sierra} J. Dubail, N. Read and E.H. Rezayi, Phys. Rev.~{\bf B 85}, 115321 (2012);
Phys. Rev.~{\bf B 86}, 245310 (2012);
I.D. Rodriguez and G. Sierra, Phys. Rev.~{\bf B 80}, 15303 (2009);
 L. Charles and B. Estienne, Commun. Math. Phys.~{\bf 376}, 521 (2020);
B. Estienne and J-M. St\'ephan, arXiv:1911.10125[cond-mat.str-el].

\bibitem{kar1} D. Karabali, Phys. Rev. {\bf D 102}, 025016 (2020).

\bibitem{kar2} D. Karabali, {\it Aspects of higher dimensional quantum Hall effect: Bosonization, entanglement entropy}, Corfu 2021 proceedings.

\bibitem{nair3} V.P. Nair, Nucl. Phys.~ {\bf B750}, 289 (2006).

\bibitem{MSZ} J. Manes, R. Stora and B. Zumino, Commun. Math. Phys.~{\bf 102}, 157 (1985).

\bibitem{KNindex} D. Karabali and V.P. Nair, Phys. Rev. {\bf D 94}, 024022 (2016);
Phys. Rev. {\bf D 94}, 064057 (2016).

\bibitem{BTquant} This is an old subject with many papers, a recent comprehensive review is M. Schlichenmaier, Contemp. Math.
{\bf 583}, 257 (2012).

\bibitem{nairBT} For Berezin-Toeplitz quantization in the Landau-Hall setting, see V.P. Nair, Phys. Rev. {\bf D 102}, 025015 (2020);
see also the paper by Charles and Estienne in reference \cite{sierra}.

\bibitem{nair4} V.P. Nair, Phys. Rev. {\bf D 92}, 104009 (2015).

%%%%%%%%%%%%%%%%%%%%%%%%%%%%%%%%%%%%%%%%%%%%%%%%
\end{thebibliography}
\end{document}